# Comparative Analysis of Routing Protocols for Under Water Wireless Sensor Networks

S. Ahmed[1], I. U. Khan[1], M. B. Rasheed[1], M. Ilahi[1], R. D. Khan[2], S. H. Bouk[1], N. Javaid[1]

[1]COMSATS Institute of Information Technology, Islamabad, Pakistan.
[2]COMSATS Institute of Information Technology, Wah Cant, Pakistan.

**ABSTRACT**

Underwater Wireless Sensor Networks are significantly different from terrestrial sensor networks due to peculiar characteristics of low bandwidth, high latency, limited energy, node float mobility and high error probability. These features bring many challenges to the network protocol design of UWSNs. Several routing protocols have been developed in recent years for these networks. One of the major difficulties in comparison and validation of the performance of these proposals is the lack of a common standard to model the acoustic propagation in the harsh underwater environment. In this paper we analyze the evolution of certain underwater routing protocols like VBF, DBR, H2-DAB, QELAR etc. in terms of their localization techniques, energy minimization characteristics and holding time calculations. The design of each protocol follows certain goals i.e. reduction of energy consumption, improvement of communication latency, achievement of robustness and scalability etc. This paper examines the main approaches and challenges in the design and implementation of underwater sensor networks. The detailed descriptions of the selected protocols contribute in understanding the direction of the current research on routing layer in UWSN.
**KEYWORDS :** Underwater Wireless Sensor Networks, Bandwidth, Latency, Localization, Holding Time, Robustness, Scalability.

## I. INTRODUCTION

Underwater Wireless Sensor Networks (UWSN) provide a promising solution for discovering aqueous environment efficiently for military, emergency and commercial purposes. Unmanned or Autonomous Underwater Vehicles (UUVs, AUVs), equipped with underwater sensors, are also envisioned to find application in exploration of natural undersea resources and gathering of scientific data in collaborative monitoring missions.

The underwater environment is much different from terrestrial and a number of issues need to be addressed while using sensor networks as an effective technology for underwater systems. Due to the high dense salty water, electromagnetic and optical signals cannot be transmitted for long distances in ocean because of scattering, high attenuation and absorption effect. Acoustic communication can be used to overcome this problem which provides a better means of data transfer in such an environment. Hence, available propagation speed is shifted from the speed of light to speed of sound which is five orders of magnitude slower i.e 1500 m/sec, which brings long propagation latency and end-to-end delay. Available bandwidth is severely limited (i.e. <100 kHz). Sensor nodes are generally considered as static but underwater sensors can move upto 1 to 3 m/sec due to underwater activities. Also, underwater nodes are larger in size so they consume more power and replacement of nodes or batteries is not so easy. Underwater applications require multi-hop networks where nodes transmit data to one of more sinks located at the surface level. Sinks then forward the received information to onshore control stations via RF transmissions.

*****Corresponding Author:** N. Javaid, COMSATS Institute of IT, Islamabad, www.njavaid.com.

The routing protocols that require higher bandwidth result in large end-to-end delays and are not suitable for these environments. Some of the challenges in under water communication are propagation delay, high bit error rate and limited bandwidth.

Due to the unique challenges of underwater environment, the communication protocols proposed for terrestrial networks cannot be directly applied to UWSNs. Many protocols have been proposed for UWSNs taking into account the unique features of underwater networks, including media access control, network and transport protocols. The routing protocols for UWSNs can be classified into localization-based and localization-free routing protocols. The routing protocols can take advantage of the localization of sensor nodes; however, the localization is not perfect because of the mobility of sensor nodes, and harsh environment. Rather localization-free routing protocols are highly demanded by research communities.

Recently, many routing protocols have been proposed for UWSNs. In this survey, we present some well-known routing protocols proposed for UWSNs, which can be broadly classified into two sections, localization-based and localization-free routing protocols.

- **LOCALIZATION-BASED ROUTING PROTOCOLS**

These routing protocols are based on the assumption of the localization of sensor nodes in UWSNs. In [1], the vector-based forwarding (VBF) protocol was proposed, in which a source node computes a vector from itself towards the sink and the neighboring nodes, around the computed vector up (called routing pipe), participate in forwarding the data packets. However, VBF has certain limitations, of hard assumption of localization of sensors and the unavailability of sensor nodes in the routing pipe.

Hop-by-hop vector-based forwarding (HHVBF) [13] is a successor of VBF and it employs the technique of computing the routing vector at each hop starting from each sender towards the sink. The recomputation at each hop reduces the effect of sparse density but inherits the assumption of the localization.

In [14], focused beam routing (FBR) utilizes different transmission power levels (i.e. ranging from P1 to PN) during the selection of next relay node, by broadcasting an ready to send (RTS) packet, and the receiving nodes reply with a clear to send (CTS) packet. The limitation of the FBR protocol lies in the use of RTS/CTS during the forwarding of the data packets causing increased delay and excessive energy consumption.

In [15], directional flooding-based routing (DFR) uses scoped flooding where a limited number of nodes are allowed to participate in forwarding data. The flooding zone is decided based on the angle among the source, current forwarder and the sink node, and the link quality of the neighboring nodes. DFR tries to limit the number of forwarding nodes. However, redundant packet's transmission cannot be avoided and the localization assumption limits its applicability.

- **LOCALIZATION-FREE ROUTING PROTOCOLS**

An overview of the routing protocols that do not assume any kind of localization are also presented. In [4], a novel routing protocol called depth-based routing (DBR) uses the depth of the sensor nodes as a routing metric and assumes that each node has a depth sensor. DBR suffers from redundant packet transmissions and excessive energy consumptions, because of the long propagation delay in UWSNs. In H2-DAB [7], hop-by-hop dynamic addressing-based routing protocol, the routing is performed based on an address (called HopID) assigned to each sensor node, based on the hop count from the sink node. The sink node broadcasts a Hello packet. The receiving nodes are assigned a HopID. These nodes then rebroadcast the Hello packet after an increment of one in the HopID. However, only the hop count value for the selection of the next hop node is not suitable in stringent

UWS network. In addition, the use of inquiry request and inquiry reply augments the already long end-to-end delay and consumes extra energy.

All these routing protocols [1] to [15] are compared on the basis of their localization techniques, mechanisms for energy minimization and holding time calculations, and a comparative study is conducted to evaluate their performances in different scenarios which can be quite helpful in the design of an efficient routing protocol.

## II. RELATED WORK AND MOTIVATION

Design of communication protocols for UWSNs is quite challenging because of the harsh underwater environment. Major challenges in their design are long propagation delay, high error rate, low bandwidth and energy. This research has considered these challenges along-with the different protocols designed for routing for underwater network since the advent of localization algorithm VBF [1] in 2006 which is in principle considered as the first of its type, till QELAR [12], an energy-efficient, and lifetime-aware routing protocol of 2012.

Localization algorithms usually use the geographic information of neighboring nodes to estimate the location. Because their design closely depends on factors like system deployment, available resources, accuracy requirements, etc., each algorithm almost aims for the specific application with its own advantages and disadvantages. This implies that there is such algorithm so far which is applicable across the spectrum. Hence, the application properties and requirements should be sufficiently investigated before their design. The primary goal of all localization algorithms is to make reasonable balance the performance and the various constraints. Even though the underwater and terrestrial routing techniques have some in common, they are extremely different due to challenges like poor link quality, high bit error rate, long latency, limited bandwidth, low data rate, etc.

In this research, we have considered many established underwater localization and routing algorithms to provide a great number of references for the UWSN routing protocol design. The localization schemes are divided into three categories in this report according to the sensor mobility:

Stationary network: In such a network, all nodes are static, which is an ideal scenario in the underwater environment because the underwater sensors are certainly being pushed due to the ocean current, shipping activities, etc. However, it is the fundamental for the other two networks.

Mobile network: Generally, mobile networks can be further divided into three types: unknown nodes are static, while beacons are moving; unknown nodes are moving, while beacons are static; both unknown nodes and beacons are moving. In this report, the mobile network mostly represents the second one, and the third one is called mobile swarm.

Mobile swarm: It is a more complicated scheme in which the ocean environment, the beacons and sensors both have the motion capabilities. The beacons can also be self-localized. Because the unknown nodes and underwater vehicles cooperate with each other by communication, the range and locations can be determined during this process without extra consumption.

The propagation delay is very high in UWSNs, hence selecting the shortest path towards the sink node is very important. Using physical distance metric instead of geographic location information avoids the need of localization of sensor nodes that is required by most of the protocols designed for UWSNs. Although the localization-free routing protocols are presented, they still have certain limitations. For instance, DBR [4] uses the depth information, where there is no guarantee that the next forwarding nodes can lead the packet in the right way towards the sink. Voids might be encountered when nodes with lower depths do not exist. To handle such a problem, deploying a number of sink nodes on the surface is needed in DBR. Another localization-free routing protocol, H2-DAB [7], uses a hop count metric for the selection of the next forwarding node along a path/route.

The node having a small hop count value is selected, may not be the shortest because any two neighbors may form long-distance links. The shortest path can be obtained even from a node having higher number of hop count value, if their physical hop distances are smaller.

In UWSNs, the transmission requires much energy than receiving. Therefore, reducing the number of transmissions is useful in reducing the energy consumption. Furthermore, network lifetime is also an important issue. Sensor nodes have limited energy and the replacement of batteries of underwater sensor nodes is very expensive in terms of both time and cost. Utilizing some nodes very frequently than others results in discharge of batteries of frequently used nodes (e.g. as in case of DBR [4] and H2-DAB [7]). This creates routing holes in the network and affects network lifetime. Hence, sensor nodes should perform energy balancing to improve network lifetime, and consume their energy evenly.

Although the experiment and test of UWSN is more difficult than the terrestrial one, many systems are implemented in practice, due to the increasingly demand of the underwater monitoring and exploring systems. In this report, a survey is conducted to summarize the challenges, the state-of-the-art algorithms and systems of underwater localization using WSN. Although the WSN based underwater localization has been investigated for years, there are still many challenging problems to be addressed, especially for the localization of the mobile networks and the mobile swarm.

### III. ENERGY MINIMIZATION

In this section we analyze the different routing protocols in terms of their energy efficiency. In VBF [1] protocol, the authors have developed a distributed and localized self-adaptation algorithm for the reduction of energy consumption by discarding low benefit packets. All the nodes close to the routing vector are qualified as relays. However, VBF may involve too many nodes in data forwarding in dense environment, increasing energy consumption. It is desirable to adjust the forwarding policy based on the node density, and the self-adaptation allows each node to estimate the neighborhood density (based on local information) and forward packets adaptively.

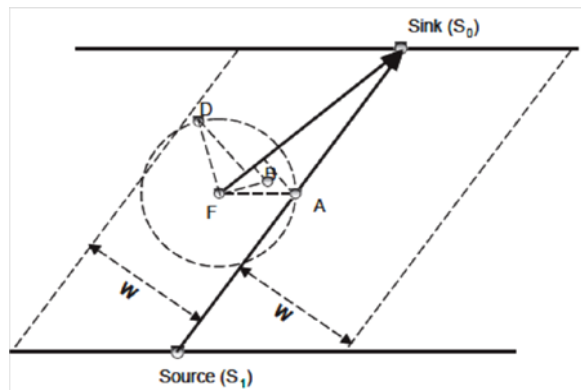

Fig .1. VBF with Self-adaptation [1]

Figure 1 illustrates the mechanism of self-adaptation in which the routing vector $S_1S_0$ specifies the forwarding path from source $S_1$ to sink $S_0$. Node F is the current forwarder. There are three nodes A, B and D in its transmission range. A has the smallest desirableness factor $\alpha$ among these, hence the shortest delay time and transmits first. Node B is most likely to discard the packet because it is in the transmission range of A and has to re-evaluate the benefit of sending the packet. Node D is out of the transmission range of A; therefore, it also forwards the packet.

If a packet receiving node finds itself close enough to the routing vector, it holds the packet for a time period Tadaptation related to α, computed as follows:

$$T_{adaptation} = \sqrt{\alpha} \times T_{delay} + \frac{R-d}{v_0} \qquad (1)$$

Where $T_{delay}$ is a pre-defined maximum delay, $v_0$ is the propagation speed of acoustic signals in water, and d is the distance between this node and the relay node.

In VBF, only the nodes close to the routing vector are involved in data forwarding, allows a node to estimate its importance in its neighborhood and adjust forwarding policy for energy saving.

DBR [4] has the advantage of handling networks with good energy dynamics. Multiple neighboring nodes of a relay may qualify to forward a packet at the next hop. It may result in high collision and high energy consumption. Hence, number of forwarding nodes needs to be controlled. To improve efficiency, a node ideally needs to send the same packet only once. To handle both these issues, DBR proposes the idea of a priority queue Q1 and packet history buffer Q2.

Q2 contains an item consisting of a unique packet ID, composed of Sender ID and Packet Sequence Number. After successful transmission of a packet, the node inserts the unique ID of the packet into Q2. If Q2 is found full, new item will replace the Least Recently Accessed (LRA) item. An item in Q1 consists of a packet and its scheduled sending time. When a node receives a packet, it holds the packet for a certain holding time. An incoming packet is inserted into Q1 if it has not been sent by the node before and it was sent from a larger depth node (i.e. dp > dc). If a packet currently in Q1 is received again during the holding time, it will be removed from Q1 if the new copy is from a node with a smaller depth (dp ≤ dc), or its scheduled time will be updated if new copy is from a lower node (dp > dc). After a node transmits a packet, it is removed from Q1 and its unique ID inserted into Q2.

The protocol introduces a parameter δ to decide the holding time of packets at each node. Each node will have a longer holding time if δ is small and end-to-end delay will be increased; lesser nodes will forward the same packet, resulting in reduction of energy consumption. Through simulations, DBR was compared with VBF, and found that the energy consumption of DBR is about half that of VBF, due to its two-queue redundant packet suppression mechanism.

FBR [14] is a scalable routing technique based on location information, and optimized for minimum energy per bit consumption. The protocol tries to reduce the unnecessary broadcast or multicast queries causing flooding and hence increases the throughput. The knowledge of location also helps to eliminate this effect. In FBR, a source node is aware just of its own location and that of final destination, but not of other nodes.

FBR algorithm can be coupled with any MAC protocol. After a multicast RTS, the requesting node may receive no answers, if there are no free neighbors. In such a situation, the transmitter will increase the transmission power, disturbing other transmissions. To handle this, a node overhearing a multicast RTS will send a short silence packet to the requesting node. Such a node will stop its transmission, minimizing the chances of interference.

Each relay node expects an acknowledgement from the current receiver. If it overhears its own packet being transmitted to the next relay, the transmitter can deduce that its last data transaction is completed. If the power level required to reach the next sensor is lower than that for the previous transmission, acknowledgement should be sent explicitly using a higher power level. Same is required when the packet reaches its final destination. If a node receives a RTS from the same transmitter for a packet that has been successfully transmitted, an acknowledgement is sent, to avoid long data packet retransmission.

Each node uses discrete four uniformly separated power control levels. System frequency allocation and bandwidth is optimized for performance in terms of energy per bit consumption, end-to-end delay and number of collisions. Average energy per bit consumption cosiders energy invested in transmission, listening and active reception of control and data packets, along-with their possible retransmissions.

E-PURLP [8] is an energy optimized routing protocol for UWSNs, consisting of a layering and a communication phase. Layering phase is responsible for minimization of energy by the technique of nodes occupying different layers. These layers are in form of concentric circles around a sink. In a particular layer, transmit energy levels are chosen such that communication occurs only with nodes in the immediate lower layer.

For layer formation, a probe of energy $E_{p1}$ is initiated at the sink node and those nodes will receive the probe whose energy is at least equal to $E_D$ (the detection threshold), and assign themselves as layer 1. Layer 1 nodes communicate with the sink in single hop. After waiting a specified time, a node of layer 1 transmits a probe of energy $E_{p2}$ to create layer 2, consisting of nodes which receive packets with energy at least equal to $E_D$ from layer 1. Waiting time is dependent only on received energy to minimize collisions between probing packets. All nodes in a particular layer can forward data to the sink over an equal number of hops. The probing energy for nodes in layer $l-1$ is related to layer width al of layer $l$ as follows:

$$E_D = \frac{E_{pl}}{a_l^{B/10} 10^{(\alpha a_l + \beta)/10}} \quad (2)$$

Layer width calculation fixes the probing energy value. By simulations, E-PULRP is found to be simple, efficient and easily implementable for UWSNs, in the absence of routing tables, localization and synchronization techniques. Only a single relay is assumed in each layer, to avoid flooding. Increasing the number of relays would increase channel contention, and a more complicated design would have to be used to avoid collisions.

H2-DAB [7] is a scalable and energy efficient novel routing protocol which uses a multi-sink architecture. In the first phase of this protocol, route creations are done by assigning dynamic HopIDs to every node in the network. In the second or last phase, data packets are forwarded towards the surface sinks by using these HopIDs.

H2-DAB formulates the analytical model for energy consumption by considering N sensors deployed uniformly in layers in an area A, from surface to bottom. Each node has an initial energy of ε, $E_d$ is the complete energy required for forwarding a packet from one layer to the other, which includes $e_d$, the energy consumed for sending data and $e_c$, energy consumed for sending the control packet. Both control packets (Inquiry Request and Inquiry Reply) are of same size and consume very little energy. Only the nodes with smaller HopID will send the Inquiry Reply.

The technique divides the depth into m layers, with each layer of n nodes and a total of D data packets generated, such that each node generates (D/N = k) data packets. Energy consumption at ith layer is $E_i$ and life time of this layer is $T_i$, while $T_i/n$ is the life time of each sensor node. All the layers can receive data packets from the below layers and forward these as well as their own generated data packets towards upper layers. HopIDs are already assigned as required only once for long intervals.

The authors have checked the energy consumption in both scenarios, static as well as mobile nodes. Every node in static scenario will send only one Inquiry Request and will get also single Inquiry Reply. Node-ID of replying node is saved in routing table to be used as a next hop. Energy consumption for a single data packet from any lower layer to next upper layer is

$$E_d = 2e_c + e_d \quad (3)$$

where "$e_c + e_d$" is the consumption from current layer which has data packet. It sends an Inquiry Request and forwards the packet after receiving the Inquiry Reply. Remaining "$e_c$" is the consumption from upper layer when a node replied with the Inquiry Reply.

For mobile nodes, the equation for $E_d$ becomes in worst case as

$$E_d = (n+3) e_c + e_d \qquad (4)$$

where "$e_d + 3e_c$" is the energy consumption from current layer, for the worst case when it has to make three Inquiry Requests. It may happen that no node replies in first two tries and then after the 3rd request, all nodes have replied from the upper layer. In such case, "$n.e_c$" will be the consumption in the form of Inquiry Replies.

Upper layers face more energy consumption problem as the number of layers starts to increase in the network. For single sink architecture, only a few nodes around the sink process all the data generated, while this burden is shared by the whole upper layer instead of few nodes in case of multi-sink architecture. To reduce this effect, Courier nodes are introduced for collection of data packets directly from the lower layers, so that upper layers process less data, increasing life of the network. The algorithm provides better results than DBR and with different parameters. DBR faces problems when nodes start to increase then energy consumption is high and when nodes start to decrease then delivery ratios are affected. Comparatively, H2-DAB maintains good delivery ratios with small number of nodes and improves with controlled energy consumption when nodes start to increase.

CARP [9] is a channel-aware cross-layer routing protocol based on a handshake mechanism for joint channel access and relay selection, and correct exchange of control packets. Once a neighbor is selected as relay, the channel is reserved and used for data transmission. An acceptable PER (packet error rate) for short control packets might result in a high PER for data packets, as they are longer. CARP is designed to obtain desirable PER for both control and data packets. Power used to transmit PING packets is computed to obtain a PER corresponding to a given BER. Once a relay is selected, power for sending data is increased so that the corresponding PER is the same as experienced by the PING/PONG exchange.

Transmission power P for transmitting packets at a given PER is computed by using a BPSK modulation, the probability of transmission of a packet $l$ bits long correctly is $(1 - BER)^l$. The BER is computed as $(1/2)$ erfc($\sqrt{SNR}$), where erfc() is the complementary error function. The SNR is given by $\{P/A(r,f)\}/\{N(f)\Delta f\}$, where P is the required transmission power, $A(r, f)$ is the attenuation in the underwater channel over a distance r for a signal of frequency f, $N(f)$ is the noise power spectral density, and $\Delta f$ is the receiver noise bandwidth.

CARP outperforms FBR and DBR, due to its link quality-based relay selection and data relaying on links that are robust for both control and data packets. DBR consumes more than CARP and FBR for delivering a bit, because being a flooding-based protocol, it incurs a higher number of data packet transmissions. As it correctly delivers a lower number of bits to sink and each bit travels longer routes than those of CARP and FBR, its energy demands are higher.

EEDBR [3] is an energy-efficient routing protocol, which utilizes the depth of sensors for forwarding data, and the residual energy to improve network lifetime. EEDBR consists of two phases: knowledge acquisition and data forwarding. During the knowledge acquisition phase, sensors share their residual energy and depth information among neighbors. In data forwarding, packets are transmitted from the nodes to the sink.

In underwater sensor networks, suppressions of packet transmissions reduce energy consumption and hence improve energy. However, too much suppression of packet transmissions affects the delivery ratio. To have a trade-off between these two parameters, authors have employed an application-based suppression scheme, such that when the delivery ratio is less than a given threshold, the number of nodes is reduced to meet the desired delivery ratio. During forwarding phase, the source includes the number of packets generated by that source. Upon reception, the sink node computes the delivery ratio by dividing the number of data packets received at the sink to the number of data packets generated by the source. If the delivery ratio is less than desired, the sink informs the source by sending a packet containing the delivery ratio. The relay node then decides whether to suppress or transmit the packet. The forwarding nodes generate a random number. If the random number is less than the delivery ratio, the packet is transmitted without any suppression, and the degree of packet transmissions is controlled. There is a tradeoff between the energy efficiency and the delivery ratio, and EEDBR, can be switched interchangeably based on the application.

Energy consumption of DBR is higher than EEDBR due to excessive number of nodes involved in forwarding the data packet and redundant transmissions in DBR. In both the schemes, the energy consumption increases with the increase in network density, as more sensors become eligible for relaying. However, DBR only restricts the number of nodes on the basis of the depth. In contrast, EEDBR restricts the number of nodes, based on the depth as well as the residual energy. Also, in EEDBR, nodes have enough difference in their holding times due to priority assignment.

RROCH [6] is an energy constrained routing protocol, to minimize the power consumption and improve reliability of data transmission. Instead of transmitting data from source to sink directly, the authors have suggested a clustering technique which leads to better energy consumption in underwater sensor environment.

The function of this protocol is based on iterations like LEACH. Each iteration begins with a initialization phase when cluster heads are selected and clusters organized, followed by a data transfer phase when the intra-cluster information is exchanged, the member nodes are chosen, and merged data transmitted to inter cluster heads. In under water networks, the total energy consumed is the sum of transmitter Energy, Receiver Energy, Sensing Energy and Computation Energy, given as,

$$E_{total} = E_{tx} + E_{rx} + E_S + E_C \qquad (5)$$

To transmit a data packet from one node to another over a distance d, the energy dissipation in underwater channel of each node is

$$E(d) = E_t(d) + E_r(d) \qquad (6)$$

$$E_t(d) = l(E_{elec} + \varepsilon_{amp}) + P_t \times \frac{l}{\alpha \times B(d)} \qquad (7)$$

$$E_r(d) = l(E_{elec} + E_{DA}) + P_r \times \frac{l}{\alpha \times B(d)} \qquad (8)$$

where $P_t$ and $P_r$ are the transmit and receive powers respectively, dependent only on the complexity of the receive operations, $l$ is packet size; $B(d)$ is the bandwidth available and $\alpha$ is the bandwidth efficiency of the modulation in bps/Hz, $E_{elec}$ is the unit energy consumed to process one bit of message, $\varepsilon_{amp}$ is energy consumed by amplifier and $E_{DA}$ is the energy for data aggregation.

The energy consumed by the cluster head is given by equation

$$E_k = N_{asn}\left(E_{elec} + \eta\big(E_{elec}T_b E_t \varepsilon_{amp}\big)\right) + l\varepsilon_{amp}d_{snk}^2 \qquad (9)$$

where $N_{asn}$ is the number of associate nodes, $\eta$ refers to data aggregation ratio, $T_b$ is bit duration.

The energy consumed by the non-cluster head nodes is given by equation

$$E_{ki} = lE_{elec} + T_b P_t \varepsilon_{amp} \tag{10}$$

The total energy consumed by $k$ clusters is therefore given by

$$E_{total} = \sum_1^N E_k + \sum_1^N E_{ki} \tag{11}$$

or

$$E_{total} = \left[ N_{asn} \left( E_{elec} + \eta(E_{elec} T_b E_t \varepsilon_{amp}) \right) + l\varepsilon_{amp} d_{snk}^4 \right] + lE_{elec} + T_b P_t \varepsilon_{amp} \tag{12}$$

RROCH and LEACH protocol has significant decrease in average energy consumption when subjected to high traffic conditions because of increased route discovery messages. RROCH has higher delivery ratio and throughput for lesser number of connections than LEACH but LEACH performs better at high traffic conditions.

A novel routing algorithm for UWSNs was designed by the name of TSR [10] (time-slot based routing), to reduce energy consumption and extend network lifetime. A probability balanced mechanism is then devised and applied to TSR. Theory of network coding is applied to meet the requirements.

In the basic TSR establishment, sink sends the broadcast packet during the first period in its own time slot. Each node which received the broadcast packet sends back a feedback packet in a particular time slot. If the sink receives the feedback, the sink will register these nodes, as first layer child-nodes. When a node first receives a packet which is not from the sink, it would determine the current time slot and whether it had missed its own time slot. If not, it will broadcast the packet in its own time slot and if yes, it would wait for another own time slot. If node x received a packet from other node y, it registers y as its sub-layer. Then, it sends the packet to parent-node z. Node z also sends a packet to its parent-node a, until the parent-node is the sink. This process will continue until all nodes are registered. The interior communication process through a route tree would decrease conflict probability efficiently and decrease the establishment time of a route tree.

In TSR, each underwater node has to maintain two tables: a state table and a table of destinations. The state table contains residual energy, total data size of sent packets, throughout capacity, error rates, average delay and repeat send times. The table of destinations contains the addresses of destinations, next hop addresses, and time messages created by the route.

In order to understand the impact of channel contention and the interaction of multiple flows in the networks, the underwater acoustic channel model had been implemented in NS-2. The research uses the MAC protocol based on TDMA, and divides each period to nine time slots, and in each time slot, only one node can send or receive a data frame. The simulation results of the protocol highlight that it can reduce the probability of node conflicts, balance energy consumption of each node, shorten the process of routing construction and effectively prolong the network lifetime.

R-ERP$^2$R [11] is again a reliable energy-efficient routing protocol. The main idea behind this protocol is to utilize physical distance as a routing metric and balance energy consumption among sensors. Also, during the selection of forwarding nodes, link quality towards the forwarding nodes is also considered to provide reliability and residual energy of the forwarding nodes to prolong network lifetime. During the data forwarding, a node that is closer to the sink than the sender, having high residual energy and having good link quality can be selected as a next forwarding node. R-ERP$^2$R attempts to avoid redundant packet transmissions and also tries to improve the delivery ratio by considering links quality.

Suppose node i is the sender of the data packet, and node j is a candidate forwarding node. Then, the link cost between nodes i and j is computed as

$$\boldsymbol{Cost\ (i,j) = (1 - RE.(j)\ /REmax) + (1 - ETX\ (i,j)\ /ETXmax)} \tag{13}$$

where RE(j) is the discrete value of the residual energy of node j, REmax is the initial/total energy of a node. ETX (i,j) is the computed ETX value of the link between nodes i and j , and ETXmax is the maximum value of the ETX, a system parameter set according to the environment.

In the start of this protocol, an initialization phase is activated, where the sensor nodes compute physical distance and expected transmission count (ETX) values and share their residual energy information among their neighbors. Then, in data forwarding phase, forwarding nodes are selected based on a cost based on ETX and residual energy, and data packets are forwarded from each source to the sink. A cost updating and maintenance phase is performed periodically to update the physical distance, ETX values and residual energy information.

The energy consumption of DBR increases continuously with the increase in the number of nodes. Hence, more nodes are involved in forwarding, increasing the overall energy consumption. In comparison, ERP²R has lower energy consumption than DBR, because ERP²R allows a limited number of nodes to forward the data packets. In ERP²R, the node having the highest residual energy (i.e. highest priority node) among its neighbors has zero holding time, while in all other candidates, forwarding nodes hold the data packet for a certain time. Hence, the highest priority node forwards the packet as soon as it receives it. Upon overhearing the data packet transmitted by the highest priority node, all other candidates forwarding nodes drop the same packet. However, in a case where some candidates forwarding nodes do not overhear the forwarding of the highest priority node, they also forward the same packet. Therefore, multiple transmissions are unavoidable completely. In contrast, in R-ERP²R, only a single node is allowed to forward the packet. Hence, the redundant packet transmissions are avoided, resulting in lower energy consumption. In addition, in R-ERP²R, the increase in the number of nodes does not have much effect on energy consumption, because only a single forwarding node is allowed to forward in all topologies.

QELAR [12] is a machine-learning-based adaptive routing protocol for energy-efficient sensor networks. It determines the behavior and performance of the agent and the goal of employing Q-learning is to get the packet delivered with minimum cost.

With the constraints that each packet forwarding attempt consumes energy, occupies channel bandwidth, and contributes to the delay, the agent is compelled to choose the relatively shorter paths to the destination, and routing delay is minimized. In a network, farther an intermediate node is from destination in terms of hops, the more negative reward it would receive, and its V value $V(s_n)$ = maxa $Q(s_n,a)$ is lower. As the greedy Q-learning algorithm always chooses the highest Q, the V value leads packets to be relayed from source to sink with minimum hops.

C(sn) is a cost function of residual energy of node n, defined as

$$\boldsymbol{c(s_n) = 1 - \frac{E_{res}(s_n)}{E_{init}(s_n)}} \tag{14}$$

where $E_{res}(s_n)$ and $E_{init}(s_n)$ are the residual energy and initial energy of node n, respectively. With initial energy Einit(sn) to be same for all sensors, the less residual energy node n has, the higher cost $c(s_n)$ is.

The reward of energy distribution in the group is defined as:

$$d(s_n) = \frac{2}{\pi}\arctan(E_{res}(s_n) - \overline{E}(s_n)) \qquad (15)$$

Larger the difference between residual energy of a node and its group average, more advantageous to be chosen as next forwarder. By definition, both $c(s_n)$ and $d(s_n)$ are in the range of [-1, 1], which enables us to balance all the parameters.

Although in Q-learning-based routing protocol, each node has to carry out some computations frequently, the computations are simple and their delay and power consumption are much smaller than that of acoustic communications. Hence, the computation overhead is ignored.

Because less energy is consumed in QELAR and the residual energy is distributed more uniformly, the most frequently used node lasts longer in QELAR than in VBF, which leads to a longer lifetime defined by the death of the first node. In general, QELAR achieves average 20 percent longer lifetime than VBF.

| PROTOCOL | TECHNIQUE | DEPENDENT PARAMETERS | OBJECTIVE |
|---|---|---|---|
| **VBF** | Localized and distributed self-adaptation algorithm | Avg distance among nodes d; Desirableness factor α; Pre-defined max delay $T_{delay}$; Transmission range R | Forwarding policy based on node density |
| **DBR** | Controlling number of forwarding nodes | Priority queue Q1; Packet history buffer Q2; Parameter δ; | lesser nodes forwarding the same packet for reduction of energy consumption |
| **FBR** | Scalable routing technique using location information and CTS/RTS | Location of source node; Location of destination; | System frequency allocation and bandwidth optimization lessening energy per bit consumption |
| **E-PULRP** | Energy optimized path unaware layered routing | Energy detection threshold $E_D$; Layer width a; Probe layer energy $Ep_l$; | Simple, efficient and flooding avoidance |
| **H2-DAB** | Uses multi-sink architecture assigning dynamic HopIDs to sensor nodes | Inquiry requests; Inquiry replies; | Maintains good delivery ratios improving controlled energy consumption with scalability |
| **CARP** | Designed to take advantage of power control for desirable PER for control and data packets | Complementary error function erfc(); transmission power P; A(r, f) attenuation in the channel; noise power spectral density N(f); Δf receiver noise bandwidth; | Link quality-based relay selection and data forwarded on robust links for control and data packets |
| **EEDBR** | Utilizes the depth of nodes for forwarding data packets, and residual energy to improve network lifetime | Current energy; Initial energy; Max holding time; Priority value p; | Suppressions of packet transmissions for reduction of energy consumption |
| **RROCH** | Clustering technique leading to better energy consumption | $P_t$ and $P_r$ transmit and receive powers; $l$ packet size; B(d) bandwidth; α bandwidth efficiency in bps/Hz; $E_{elec}$ unit energy consumed to process one bit; $\varepsilon_{amp}$ energy consumed by amplifier; $E_{DA}$ energy for data aggregation. | Minimizing power consumption and improving reliability of data transmission. |
| **TSR** | Time-slot based routing and probability balanced | A state table; A table of destinations. | Reducing node conflicts, shortens the routing construction, balances |

| | mechanism | | energy consumption of nodes and prolongs the lifetime |
|---|---|---|---|
| **R-ERP$^2$R** | Utilizes physical distance as a routing metric and balances energy consumption among sensors | RE(j) the residual energy of node j; REmax the initial/total energy of a node; ETX (i,j) the ETX value of the link between nodes i and j; ETXmax is the maximum value of the ETX | Avoid redundant packet transmissions resulting in lower energy consumption. |
| **QELAR** | Machine-learning-based adaptive routing protocol for energy-efficient sensor networks | C($s_n$) a cost function of residual energy of node n; reward of energy distribution in the group d($s_n$) | Each node carries out simple computations; delay and power consumption are much smaller, ignoring computation overhead |

Table 1*: Energy MinimizationTechniques*

## IV.    LOCALIZATION

Several techniques have been proposed for the optimization of routing issues in under-water sensor networks, but most of them are localization based. Review of some of those techniques is described below.

In this regard, the first and most adopted in future protocol came on screen in 2006 Vector-Based forwarding VBF [1] protocol for UWSNs. It is a location-based geographic routing approach aiming to provide scalable, robust and energy-efficient routing. This routing algorithm also handles node mobility efficiently, in addition to energy saving.

Each packet in VBF carries the information of the location of the sender, the target and intermediate nodes. The routing vector specifies the forwarding path from sender to target. When a node receives a packet, it measures its distance to the forwarder along-with the angle of arrival of the signal and computes its relative position. All the packet receiving nodes compute their positions in a similar way. When a node determines that this routing vector or distance is less than predefined threshold value W i.e. close enough, it forwards the packet by attaching its own computed position with it. Hence, a routing pipe is formed by all the packet forwarders in the network and all those nodes in this pipe will be eligible for packet forwarding, and remaining would not be. Fig. 2[1] below represents the basic VBF idea, where node S1 is the source, and node S0 the sink. S1S0 represents the routing vector.

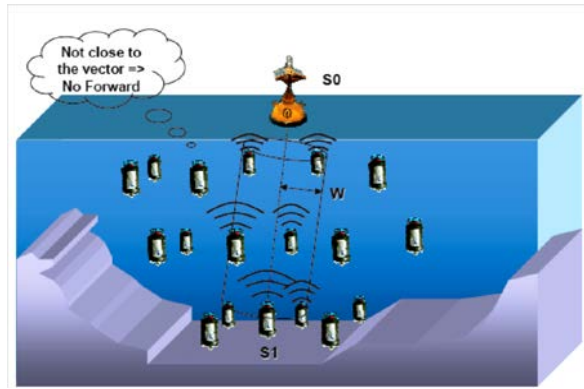

Fig .2. A general view of VBF for UWSNs

No state information is required by VBF at each node, only the nodes along the forwarding path take part in packet routing, which saves the overall network energy. In case of dense deployment, the protocol involves sufficient nodes in data forwarding, hence increasing the energy consumption.

So, the authors have introduced a factor called desirableness denoted by α which measures the stability of a node A to forward the packets to a node F, given by the expression (1):

$$\alpha = \frac{p}{W} + \frac{(R - d \times cos\theta)}{R} \tag{16}$$

where d is the distance between nodes A and F, p is A's projection on the routing vector S1S0, and θ denotes the angle between vector FS0 and vector FA. W is the radius of the "routing pipe" or the threshold value and R is the transmission range. This numerical expression can be depicted by the given fig 3 [1] below.

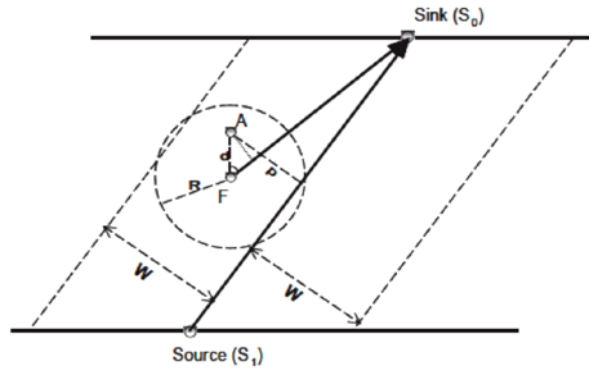

Fig .3. Calculation of Desirableness Factor

If α of a node comes out to be 0, then that node is called optimal, and is at its best position, and if its value is close to 0, then the node is close to its best location.

VBF is strong against node failure and packet loss and uses redundant paths in forwarding the packets. Some paths may be interleaved while some are parallel. All nodes are assumed to be deployed in layers, and the adjacent layers are separated by a distance of R/2. Those nodes which are present inside the cylindrical routing pipe qualify as forwarders. If d' denotes the density of nodes, $p_l$ the loss probability of packets, pe the failure probability of nodes, and h the number of layers or hops. The number of nodes in each layer is computed as $N_l = (\pi \times W^2 \times R \times d)/2$, and forwarding ones at each layer are $N_l \times (1 - p_e)$. Transmission space of a node is a sphere with radius R and having 3 layers. All those nodes which lie inside the sphere will hear the transmission of the packet if done by any node and the number of nodes in each layer are $n_t = (4/3) \times \pi R3 \times d' \times (1/3)$.

The probability of reception of any packet by the nodes in the upper layer is computed as

$$P_1 = 1 - (1 - (1 - p_l)(1 - p_e))^{n_t} \tag{17}$$

In 2007, another protocol was developed by the name of HH-VBF [13], Hop-by-hop Vector-based forwarding, based on the concept of routing vector of VBF.

Based on the above limitations found in VBF, this protocol defines a virtual pipe around the per-hop vector from each forwarder to the sink instead of using a single virtual pipe from source to sink. Each node then can adaptively make packet forwarding decisions based on its current location. This not only enhances data delivery ratio in sparse networks but is also less sensitive to the routing pipe radius threshold.

The authors modified equation 1 to compute the desirableness factor α' of a node A as in equation below

$$\alpha' = \frac{(R - d \times \cos\theta)}{R} \quad (18)$$

When any node in HH-VBF receives a packet, it holds it for some period of time. This time will be proportional to its desirableness factor and, the node having the smallest value of desirableness factor will be the first one to send the packet. Each node in the vicinity may hear the same packet multiple times, and calculates its distances to the various vectors from the packet forwards to the sink. This node will only forward the packet if the minimum one of these distances is still larger than a pre-defined minimum distance threshold β. Bigger the value of β, more nodes forward the packet. Forwarding redundancy can be controlled by adjusting β.

With the help of simulations, it was noted that HH-VBF and VBF are robust to node mobility, while HH-VBF has quite better performance in terms of average energy consumption and success rate in sparse networks.

Hai Yan et.al presented their protocol by the name of DBR: Depth-based Routing for Underwater sensor networks [4] in 2008. Unlike VBF and HH-VBF protocols which require complete location information of all sensor nodes, DBR only needs the local depth information and can handle network dynamics much efficiently.

DBR is a greedy algorithm and forwards data packets towards the water surface based on the depth information of each sensor. When a node receives a packet, it only forwards the packet if its depth is less than that present in the packet. The packet format in DBR is shown below in fig 4.

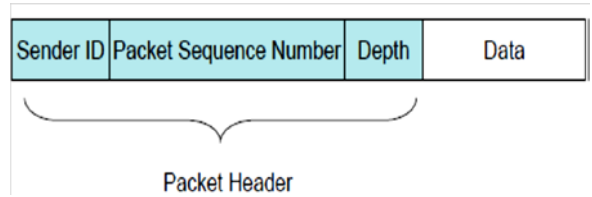

Fig .4. DBR Packet Format

Each node in DBR maintains a priority queue Q1 and a packet history buffer Q2. When a node transmits a packet, it inserts the unique ID (Sender ID plus Packet sequence number) of the packet into Q2. When Q2 is full, the new item will replace the Least Recently Accessed (LRA) item. The priority of an item in Q1 is represented by the scheduled sending time. When a node receives a packet, it holds the packet for a certain holding time. The scheduled sending time of a packet is computed based on the time when the packet is received and the holding time for the packet.

An incoming packet at a node is inserted into Q1 if its unique ID is not in Q2, and has a larger depth than others. If a packet currently in Q1 is received again during the holding time, the packet and its scheduled sending time will be updated if the new copy is from a node with a lower depth. After a node sends out a packet, it is removed from Q1 and unique ID inserted into Q2. The holding time for a packet is the difference d between depth of the packet's previous hop and that of current node.

Fig 5 below shows a pictorial representation of DBR protocol in which node S is the sender, and nodes n1, n2 and n3 are all its one-hop neighboring nodes. As n3 is below S, so it discards the packet and n1 is preferred because of its lower depth, as shown below.

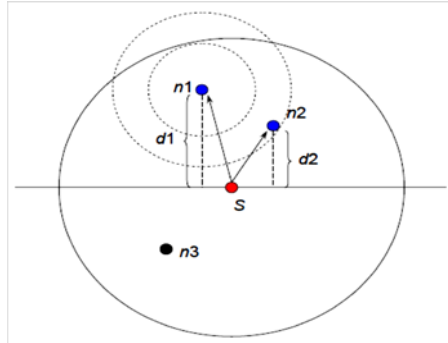

Fig 5: DBR's node selection technique

The holding time is expressed as a function of d in DBR as follows:

$$f(d) = \frac{2\tau}{\delta}.(R - d), \delta \in (0, R] \tag{19}$$

where R is the maximum transmission range, $\tau = R/v_0$ is maximal propagation delay of one hop, $v_0$ is sound propagation speed in water and $\delta = d_1 - d_2$.

A small value of δ leads to nodes with longer holding times, resulting in longer end-to-end delays. Forwarding at these nodes is likely to be suppressed than the one closer to the water surface, which results in lower energy consumption. Total energy consumption of DBR is about half that of VBF, and achieves a better delivery ratio for sparse networks. Delay in VBF is shorter than DBR in one-sink case and packets can be delivered to any sink in multi-sink case.

DBR protocol requires more memory to maintain two buffers, but as the UWSNs have relatively low data, so only small buffers will be needed.

Another world-wide acknowledged protocol also appeared in 2008 known as FBR [14]: Focused Beam Routing Protocol for Underwater Acoustic Networks by Joseph Jornet et al. This cross-layer approach of MAC and physical characteristics is suitable for networks containing both static and mobile nodes which may or may not be synchronized.

FBR is a distributed algorithm in which a route is established dynamically as the data packet moves towards its destination shown in fig 6 below.

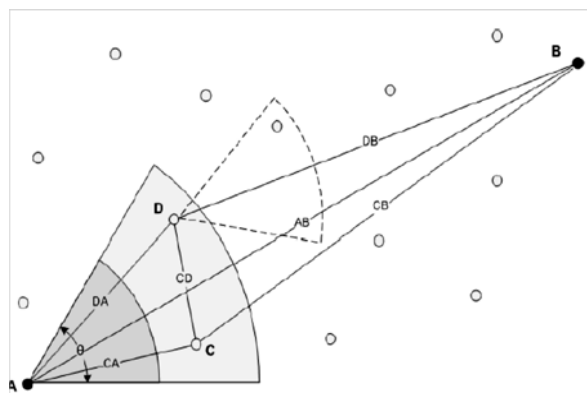

Fig .6. Illustration of FBR protocol

As in the fig, node A issues a RTS (request to send) signal as it wants to transmit a packet to node B, the packet contains the location of both these nodes. It is a multicast request and the initial transaction takes place at the lowest power level and the power is increased only if necessary from P1 through Pn. For each power level Pn there is a transmission radius dn, and the nodes within this radius receive a detectable signal. All the nodes that receive A's multicast RTS calculate their relative location from the AB line, to find if they are candidates for relaying. They will be considered as candidates if they lie within a cone of angle ±θ/2 emanating from the transmitter towards final destination. Only such a node will respond to the RTS. A transmitting node will keep raising the power level until all power levels have been exhausted. If maximal power level Pn is not reached, the transmitter will shift its cone and start looking for candidate relays on both sides of the main cone. This technique suits the paths where zigzagging is minimum, which guarantees that all possible paths will be finally found.

When RTS is received by any node, it is confirmed for relaying, and replies using a clear to send (CTS) signal. This signal has the name and location of the issuing node issuing (C or D) and the addresses of the source and destination (A and B). The two nodes replies may collide. However, as the CTS is very short, and the distances CA and DA are never the same, chances of collision are minimal. If there is no collision, A receives both replies. A knows which candidate is closer to the final destination due to sender's location. If D is chosen as the relay, then packet is transferred to it, C will overhear the data packet transaction and deduce that it has not been chosen. This avoids the risk of data packet collisions, and only packets that can collide are the short control packets.

Another technique was presented in 2008 by the name of SBR-DLP, Sector- based routing with destination location prediction for underwater mobile networks by Nitthita Chirdchoo et.al [5], which is a location-based routing protocol designed to help enhance the packet delivery rate.

SBR-DLP shares some similarities with FBR but has some considerable differences. SBR-DLP does not assume that the destination node is fixed and location accurately known, like FBR. Unlike FBR, SBR-DLP considers the entire communication circle to locate the candidate relay nodes. SBR-DLP does not need to rebroadcast RTS every time. The CTSs from different neighbors may collide, which degrades its performance. Also SBR-DLP does not assume the knowledge of all other nodes' movements and avoids flooding by routing a packet in a hop-by-hop fashion.

The working of this protocol can be illustrated by the following fig 7.

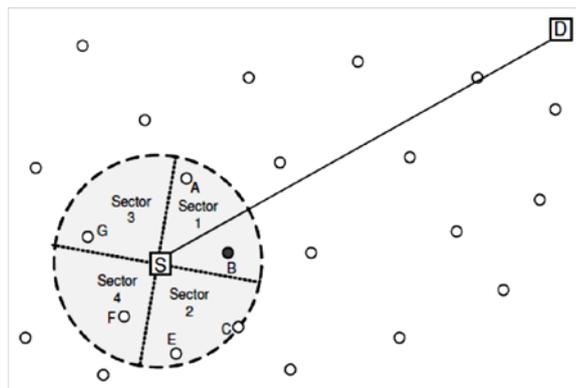

Fig .7. Forwarder Selection by SBR-DLP

A node S responds to destination D by finding its next relay node. This is done by broadcasting a Chk_Ngb packet, which includes packet ID and sender's current position. Each neighboring node then checks its closeness to node D. If node x satisfies the condition, it will respond to S by transmitting a Chk_Ngb Reply packet. To reduce collisions at Node S, each neighboring node first determines its sector, and then schedules the transmission time of its Chk Ngb Reply. Figure 7 highlights the labeling of a four-sector system. The selected node x then writes into its Chk_Ngb Reply the sector number, its node ID, and computed distance from the destination location. The transmission is scheduled to occur after an offset found as

$$toffset, j = α(j − 1)Pmax \qquad (20)$$

where α lies between 0 and 1 and depends on the number of sectors k, Pmax is the maximum propagation delay.

After all the Chk_Ngb Reply signals are received from candidate neighbors, node S filters out the out of range nodes, using its propagation delay from each candidate, time of reception of Chk_Ngb Reply, and maximum relative velocity. The remaining candidates are sorted according to their sector priorities. A tie if occurs will be broken by the closest predicted distance to destination D. Node S now transmits the data packet to this node. The relay node acts as a sender using the same procedure. If there is no response from any of the sender's neighbors, it will wait for a time interval, before another attempt. If the sender fails for n discard times, it drops the packet.

SBR-DLP is a multi-sector based routing algorithm coupled with destination location prediction and is suitable for environment where destination nodes can also move along-with other nodes. Its design considers the features of long propagation delay, high channel error rate, node mobility, and low data rate.

DFR: an efficient directional flooding-based routing protocol in underwater sensor networks [15] was proposed in 2011 by Dongseung et al. DFR performs controlled flooding in order to achieve reliable packet delivery. It also follows the techniques of VBF and HH_VBF. The protocol varies the number of nodes participating to forward a packet based on their link quality.

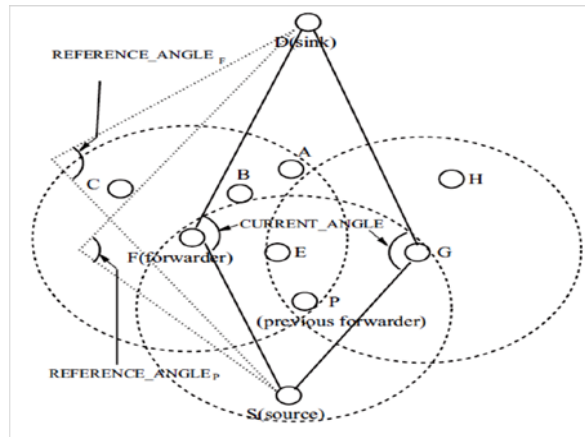

Fig 8: DFR node selection technique

Fig 8 depicts the working of DFR protocol in which a source S broadcasts a packet which consists of its location and initial REFERENCE ANGLE, set to a predefined minimum value A_MIN. If the packet arrives at a node P and is rebroadcasted, the packet includes its updated REFERENCE ANGLE value, $RA_P$. When a forwarding node F receives the packet from P, F decides its forwarding by comparing its $CA_F$ with $RA_P$ in the packet. $CA_F$ is obtained by the law of cosines:

$$CA_F = \arccos\left(\frac{|FS|^2+|FD|^2-|SD|^2}{2.|FS|.|FD|}\right) \qquad (21)$$

If $RA_P$ is greater than the $CA_F$, F drops the packet as considered out of flooding zone. F then adjusts $RA_F$ and forwards the packet. If no neighbor node is found closer to the sink than F, F executes the void handling process. $RA_F$ is adjusted based on link quality to its neighbors. If average link quality AvgLQ to neighbors is found worse than the predefined threshold LQth, F sets $RA_F$ to the value of $RA_P$−A_DCR, where A_DCR is a predefined decrement value. This allows more nodes to participate in forwarding the packet. If AvgLQ is better than LQth , F sets $RA_F$ to the value of $RA_P$ + A_ICR, where A_ICR is a predefined increment value. This causes fewer nodes to participate in forwarding the packet.

$$RA_F = \begin{cases} RA_P + A\_ICR, & if\ AvgLQ > LQ_{th} \\ RA_P, & if\ AvgLQ = LQ_{th} \\ RA_P - A\_DCR, & if\ AvgLQ < LQ_{th} \end{cases} \quad (22)$$

F sets its forwarding delay based on RAF. If every node transmits the packet simultaneously, collisions might occur and a forwarding delay can help to alleviate them.

$$FORWARDING\_DELAY = \left(\alpha\left(1 - \frac{Dist(F)-Dist(P)}{TX\_RANGE}\right) + \beta\left(1 - \frac{1}{AvgLQ}\right)\right) \times MAX\_PROPOGATION\_DELAY \quad (23)$$

In the above equation, Dist(X) shows the distance between node X and the sink, TX_RANGE is its transmission range, and MAX_PROPAGATION_DELAY is the maximum propagation delay of the acoustic wave. If α is 1 and β is 0, the forwarding delay favors the advancement towards the sink. If α is 0 and β is 1, then forwarding delay is only dependant on the link quality.

Through NS-2 simulations, it is observed that DFR performs better than VBF and HHVBF in terms of communication overhead and packet delivery ratio, considering node mobility.

2011 also saw the emergence of a very efficient protocol by the name of H2-DAB, Hop-by-Hop Dynamic Addressing Based protocol [7], to handle the problem of node mobility. Every node in the network is assigned a routable address in an efficient way without requiring any dimensional location. This helps nodes to communicate without any centralized infrastructure and they can come and leave the network without having any rest effect.

The first phase of H2-DAB creates routes by assigning dynamic HopIDs to every sensor in the network. In the second phase, data packets are forwarded towards the surface sinks by using HopIDs. HopID is used for routing decision whereas Node-ID is for node identification. Each node gets its HopID dynamically, and is variable with the node movements. Node-ID is a unique address for every node throughout its life time in the network.

Every ordinary sensor node uses a default value ''99'' as its HopID and ''0000'' as Sink-ID in routing table, till it has not received any hello packet. After reception of a hello packet from any surface sink, or ordinary node with a minimum power threshold PTmin, it starts to update its HopID. It then forwards the S-hp with its new S-HopID. The receiving nodes will increment their S-HopIDs by one, and will continue forwarding them towards their neighbors, till S-hp becomes zero.

If a source node does not get the response from its neighbors with smaller HopIDs, it will wait for a $t_1$ amount of time and try again. After the third attempt, if the result is same, it assumes that no such node is available, and it can forward the data packet towards a node on the same layer with the HopID value nearly or equal to its own HopID or lower layer nodes.

$t_1 = \frac{C}{(n_1+1)}$, where C is a constant, having the maximum value of the waiting time and $n_1$ is the number of neighbor nodes replied in the first inquiry request.

If it still cannot find any node after the 2nd try from the upper layers, it will wait $t_2$ time depending on the number of nodes replayed after the 2nd inquiry request and, the difference between the number of nodes in the 1st and 2nd inquiry request. An average of these parameters will be acquired.

$$t_2 = \frac{[\frac{c}{|n_2-n_1|+1}+\frac{c}{n_2+1}]}{2} \qquad (24)$$

Performance of this protocol is compared with that of DBR protocol and found that in H2-DAB, the delivery ratios are not based on the density or sparseness of sensor nodes. Node mobility due to water currents and node failure are the challenges handled easily with this protocol. New nodes can be added at any time and can configure easily during next interval.

Another protocol based on the backgrounds of FBR and DBR protocols was proposed by Stefano et al. in 2011 by the name of CARP, Channel-aware routing protocol for underwater wireless networks [9]. The protocol combines hop count information with link quality to route around connectivity voids and shadow zones, giving advantage of power control for robust links.

At the start-up, HELLO packets are flooded from the sink throughout the network. Sink generates the first HELLO packet, setting its hop count field to 0, and broadcasts it to its one hop neighbors. Each node x receiving an HELLO packet checks whether its HC(x) is greater than the hop count embedded in the packet plus 1. If so, x updates its hop count by plus 1, and re-transmits the packet. Otherwise, the packet is dropped. By the end, a node has acquired its hop distance from the sink, as well as information about its neighbors.

When a node x has one or more data packets to forward, it chooses a suitable relay node, by broadcasting a control packet, PING. A node y that receives the PING packet replies with a PONG packet to the source x. Node x awaits for PONG replies for a time δ. δ depends on the nominal transmission range and acoustic signal speed in water. It is continuously updated by the actual round trip time of PING/PONG handshakes. After time δ, node x uses the link quality information lqy sent in the PONG packets from all its available neighbors y, and combines it with the link quality from x to y, lqx,y. For each responding y, node x computes:

$$goodnessy = lqylqx, y. \qquad (25)$$

The node y with the highest ratio goodnessy/HC(y) is chosen as the relay, and data packets are sent directly to it.

Evaluation of CARP was done using ns-2 simulations and compared with those of FBR and DBR, and found that the protocol efficiently exploits short control messages to perform joint channel access and relay selection, with usage of link quality information in the cross layer relay selection.

Abdul wahid et al. proposed another routing protocol in 2012 called EEDBR [3] which stands for Energy-efficient Depth-based routing protocol, which utilizes the residual energy of sensor nodes to improve the network lifetime.

During the knowledge acquisition phase of EEDBR, nodes share their depth and residual energy information among their neighbors. In data forwarding, data packets are transmitted from sensors to the sink.

The forwarding nodes upon receiving the data packet, hold the packet for a certain time based on their residual energy. A sensor having more residual energy has a short holding time. The holding time (T) is computed using:

$$T = (1 - (\frac{current\ energy}{initial\ energy})) \times max\_holding\_time + p \qquad (26)$$

where p is the priority value and max_holding_time is the maximum holding time of a packet by a node. Priority value is used to have different holding times always, since the sensors might have same residual energy. This is initialized with a starting value, and is doubled with the increase in the position index of the nodes in the list. Due to different positions in the list, the nodes have different priority values.

In packet forwarding from a source to a sink of EEDBR, each node of the data packet includes a list of its neighboring nodes having smaller depths, called forwarding nodes, ordered on the basis of their residual energy values. The first node in the list upon receiving the data packet, forwards the data packet immediately without waiting. Rest of them holds the data packet for a certain time T. If during T, a forwarding node overhears the same data packet from another sensor, it generates a random number and compares it to the delivery ratio received in the packet. If the random number is less than the delivery ratio, then the transmission is suppressed, and vice versa.

Performance of EEDBR is compared with routing protocol DBR through simulations, and observed that EEDBR contributes to improvements in network lifetime, energy consumption and end-to-end delay, keeping the delivery ratio almost similar to compared routing protocol.

Abdul Wahid again proposed a routing protocol in 2012 by the name of R-ERP$^2$R, [11] Reliable Energy-efficient Routing Protocol based on residual energy and physical distance. The idea behind this is to utilize physical distance as a routing metric and to balance energy consumption among sensors. It takes into account multiple metrics like link quality, physical distance and residual energy, unlike other protocols that consider separate routing metrics.

The protocol consists of 3 phases. In the start, an initialization phase is activated, where the sensor nodes compute physical distance using Time of Arrival (ToA)/Time Difference of Arrival (TDoA) and expected transmission count (ETX) values and share their residual energy information among their neighbors.

In data forwarding phase, relay nodes are selected based on cost, and data packets are forwarded from each source to the sink, using the equation below:

$$Cost\,(i,j) = (1 - \frac{RE(j)}{REmax}) + (1 - \frac{ETX(i,j)}{ETXmax}) \qquad (27)$$

where RE(j) is the discrete value of the residual energy of node j, REmax is the initial/total energy of a node. ETX (i , j) is the computed ETX value of the link between nodes i and j, and ETXmax is the maximum value of the ETX, set according to the environment.

R-ERP$^2$R uses a combination of both implicit acknowledgment and retransmission mechanism. The implicit acknowledgment is the overhearing of the data packet transmitted by the next forwarding node. Upon overhearing the same packet, forwarding node removes the packet from its buffer. In case, the packet is not overheard within a certain period of time, the forwarding node retransmits the data packet. The retransmissions are performed for a specific number of times.

A cost updating and maintenance phase is performed periodically to update the physical distance, ETX values and residual energy information. Because of the updated residual energy information, different sensors can be selected as relay at different times, leading to energy balancing.

R-ERP$^2$R was implemented in NS-2 simulator and its performance evaluated under different scenarios using grid and random topologies against DBR, and dominates in network lifetime, energy consumption, delivery ratio and end-to-end delay.

| PROTOCOL | TECHNIQUE | DEPENDENT PARAMETERS | OBJECTIVE |
| --- | --- | --- | --- |
| **VBF** | Location-based geographic routing approach | Relative position of the receiving node; Angle of arrival AOA of signal θ; Desirableness factor α; Density of nodes d; Number of hops or layers h; Radius of routing pipe R; | Provides robust, scalable and energy-efficient routing; able to handle node mobility in an efficient way. |
| **HH-VBF** | Hop-by-hop Vector-based forwarding | Desirableness factor α'; Density of nodes d; Number of hops or layers h; Radius of routing pipe R; Angle of arrival AOA of signal θ; minimum distance threshold β; | Enhances data delivery ratio in sparse networks |
| **DBR** | Depth-based Routing and needs only local depth information | Difference in depths δ=d1 and d2; Max transmission range R; Max propagation delay τ; | Handles network dynamics much efficiently without the assistance of a localization service. |
| **FBR** | A distributed algorithm in which a route is dynamically established | RTS and CTS signals; Maximal power level Pn; cone of angle ±θ/2; | Energy-efficient multi-hop communications |
| **SBR-DLP** | Sector- based routing with destination location prediction | Chk_Ngb packet; A parameter α, 0 ≤ α ≤ 1; offset time toffset; Maximum propagation delay Pmax; | Location-based routing protocol for enhancing packet delivery rate. |
| **DFR** | Directional flooding-based routing protocol | Dist(X) distance between node X and the sink node; TX_RANGE transmission range of the node; MAX_PROPAGATION_DELAY maximum propagation delay | Controlled flooding for reliable packet delivery |
| **H2-DAB** | Every node assigned a routable address without dimensional location information | HopID (for routing decision); Node-ID (node identifier); minimum power threshold PTmin; Waiting times; | Handles the problem of node mobility |
| **CARP** | Exploits link quality information for cross layer relay determination | Control PING/PONG packets; waiting time δ; goodness; | Routes around connectivity voids and shadow zones, with power control for robust links |
| **EEDBR** | Utilizes residual energy of sensor nodes to improve network lifetime | Current energy; Initial energy; Max holding time; Priority value p; | Improvements in network lifetime, energy consumption ,end-to-end delay |
| **R-ERP$^2$R** | Utilizes physical distance as a routing metric and balances energy consumption among sensors | Time of Arrival (ToA); Time Difference of Arrival (TDoA); Expected transmission count (ETX); cost (based on ETX and residual energy); | Updated residual energy of nodes leads toforwarding nodes and improves energy balancing |

Table 2: Localization Techniques

## V. HOLDING TIME CALCULATION

VBF [1] is based on self-adaptation algorithm which introduces extra delay in data forwarding, for the purpose of differentiating the importance of nodes in the transmission range. If maximum delay Tdelay is set to a smaller value, end-to-end delay can be reduced. However, Tdelay must be set large enough due to the purpose of delay time used by VBF.

If N denotes the total nodes in the network and available space be $X \times Y \times Z$, then the average distance among nodes is given by $d = \sqrt{(\Delta x2 + \Delta y2 + \Delta z2)}$, where $\Delta x = X/N, \Delta y = Y/N$ and $\Delta z = Z/N$. If W be the radius of routing pipe and R the transmission range, then the average time for the travel of an acoustic signal between two neighbor nodes is $T = d/v0$, where v0 is the propagation speed of acoustic signals in water. The delay time Tadapation in the self-adaptation algorithm must be greater than T. Let $D = min\{W, R\}$, and $\Delta\alpha$ be the difference of the desirableness factors of these two nodes, then $\Delta\alpha \leq 2 \times d/D$, and the lower bound for Tdelay is $\sqrt{(Dd)}/\sqrt{2}xV$.

The holding time at a node for a packet is calculated based on d, the difference between packet's previous hop depth and that of the current node. According to DBR [4], the holding time must satisfy the conditions of holding time which decreases with the increase of depth d; and the difference between holding times of two neighboring nodes which must be long enough. The authors have expressed the holding time using a linear function of d as $\boldsymbol{f(d) = \alpha \cdot d + \beta}$…………..(i)

If d1 and d2 are the depth differences at nodes n1 and n2, n1 receives a packet from S at time t1, n2 receives the packet at time t2, and t12 is the propagation delay between n1 and n2, then we can have

$$\boldsymbol{\alpha \leq \frac{(t_2-t_1)-t_{12}}{d_1-d_2}, (\alpha < 0)} \tag{28}$$

For the worst conditions, choose $|\alpha| = 2\tau/(d1 - d2)$ where $\tau = R/v0$ is the maximal propagation delay of one hop and R is the maximal transmission range of a sensor node. α varies from 0 to R. If $d1 - d2 = \delta$, then $\alpha = -2\tau/\delta$. For to compute β, the equation is: $(-2\tau/\delta)R + \beta = 0$. Substituting the values of α and β in equation (i), we have

$$\boldsymbol{f(d) = (2\tau/\delta)(R - d), where\ \delta = (0, R]} \tag{29}$$

**IV.1   CLUSTER FORMATION**

In Ad-LEACH, during the establishment of network the whole area is alienated into permanent and static clusters. The shape of clusters can be square or rectangular according to the design requirement and area available. During our simulation we found almost identical results of both rectangular and square shape clusters.
Each cluster contains a separate Ad-LEACH protocol running in parallel to its neighboring clusters. The inspiration in the wake of separating the whole area into small static fields is to reduce complexity and power dissipation. Small portions of clusters are easy to manage rather than one large field of operation. In this way, the nodes also reduce the power level of their broadcast messages because they only have to cover a small portion of area from the main region.

**IV.2   CLUSTERS HEAD (CH) SELECTION**

Running a separate Ad-LEACH into all clusters means that each cluster has its own CH. DEEC is proposed in [9], which takes the heterogeneous characteristics of WSNs into consideration. This results in improved scalability and a reduced amount of battery consumption. In order to acquire more definitive solution, we choose CH selection algorithm of DEEC in our Ad-LEACH protocol

The DEEC solution chooses CHs based on their residual energy. Each node requires prerequisite knowledge of network like total energy and network life time. In DEEC, BS broadcasts the total energy of network $E_{total}$ to all nodes. The BS also estimates the value of $R$ which is network lifetime and broadcast it to all nodes. In the start of every new epoch, all nodes calculate the value of $p_i$ using the equation (5) which is taken from [9], as:

calculate the value of $p_i$ using the equation (5) which is taken from [9], as:

$$p_i = \frac{p_{opt} N_{cls} (1+a) E_i(r)}{N_{cls} + \sum_{i=1}^{N_{cls}} a_i \overline{E}(r)} \qquad (30)$$

Here $N_{cls}$ is total number of nodes present in current cluster. The value of $\overline{E}(r)$ is calculated in equation(6)[9], as:

$$\overline{E}(r) = \frac{1}{N_{cls}} E_{total} (1 - \frac{r}{R}) \qquad (31)$$

Let the energy consumed by network in each round is denoted by $E_{round}$ then the estimated value of $R$ is calculated with equation (7) [9].

$$R = \frac{E_{total}}{E_{round}} \qquad (32)$$

Now each node uses the value of $p_i$ and put it in equation number (8) to get the value of $T(s_i)$ [9]. The value of $T(s_i)$ is used by every node to decide if it is CH in current round.

$$T(s_i) = \begin{cases} \frac{p_i}{1 - p_i(r \bmod 1/n)} & \text{if } s_i \in G \\ 0 & \text{otherwise} \end{cases} \qquad (33)$$

After a node is selected as CH, it must keeps its radio receiver turned on so all client nodes to inform the CH about their existence. In order to do that Carrier Sense Multiple Access (CSMA) MAC protocol is used in this phase by all client nodes.

### IV.2.1 PROBABILITY OF HETEROGENEOUS NODES

The equation (9) [9] dictates that $p_{opt}$ is the reference value of $p_i$. In heterogeneous network the reference value of every node differ from each other according to its initial energy value.

$$p_i = p_{opt} \frac{E_i(r)}{\overline{E}(r)} \qquad (34)$$

As two level heterogeneous network is considered in this research we will use modified values of $p_{opt}$ as given in equation (10) and (11). [9]

$$p_{adv} = \frac{p_{opt}}{(1+am)} \qquad (35)$$

$$p_{nrm} = \frac{p_{opt}(1+a)}{(1+am)} \qquad (36)$$

This changes the value of $p_i$ and we get equation (12).

$$p_i = \begin{cases} \dfrac{p_{opt} E_i(r)}{(1+am)\overline{E}(r)} & \text{if} s_i = \text{normalnode} \\ \dfrac{p_{opt}(1+a)E_i(r)}{(1+am)\overline{E}(r)} & \text{if} s_i = \text{advancenode} \end{cases} \qquad (37)$$

As we considered two level heterogeneous network in our research we will use modified values of $p_{opt}$ as given in equation (10) and (11).

### IV.2.2  CLIENT SCHEDULING

The CH receives the client information from each node. The CH creates a Time Division Multiple Access (TDMA) schedule for all of its client nodes and broadcast it back to all nodes. this TDMA schedule is used by all client nodes to transmit their data towards CH node.

### IV.2.3  DATA TRANSMISSION

After all client nodes receive their TDMA slot information, the process of data transmission begins. All client nodes only communicate to CH during their assigned time slot. In order to save the energy each client node turned off its radio during unallocated timeslots. The nodes lie near to CH transmit low energy signal and as the distance increases between client node and CH the transmission energy of each node increases. Each client node chooses its own transmission energy level, based on Received Signal Strength (RSS) of the CH advertisement message. When a CH receives data from all of its client nodes, it performs some necessary signal processing techniques on this data to compress it. After compression, this data is transmitted towards BS. During this whole process the radio interface of CH remained turned on, which consumes energy. When CH transmits information towards BS, it is also high energy transmission. This leads to the fact that being a CH puts a lot of energy burden on each node. That is the main reason behind rotating CHs during whole network operation.